\documentclass[11pt,a4paper]{article}

\usepackage[english]{babel}
\usepackage[utf8]{inputenc}
\usepackage{amsfonts}
\usepackage{amsmath}
\usepackage{amssymb}
\usepackage{graphicx}
\usepackage{epsfig,epsf}
\usepackage{bm}
\usepackage{verbatim} 
\usepackage{booktabs}
\usepackage{multirow}
\usepackage{slashed}
\usepackage{xcolor}
\usepackage[font=small]{caption}
\usepackage{float}
\usepackage{blindtext}
\usepackage{placeins}
\usepackage{bbm}
\usepackage{arydshln} 
\usepackage{cite}
\usepackage{soul}
\usepackage{xfrac}
\allowdisplaybreaks

\def\linkcolor{cyan!70!black}
\usepackage[
colorlinks=true
,urlcolor=\linkcolor
,anchorcolor=\linkcolor
,citecolor=\linkcolor
,filecolor=\linkcolor
,linkcolor=\linkcolor
,menucolor=\linkcolor
,linktocpage=true
,pdfproducer=medialab
,pdfa=true
]{hyperref}

\usepackage{geometry}
\newenvironment{Eqnarray}{\arraycolsep 0.14em\begin{eqnarray}}{\end{eqnarray}}
\newcommand{\ba}{\begin{Eqnarray}}
\newcommand{\ea}{\end{Eqnarray}}
\newcommand{\be}{\begin{equation}}
\newcommand{\ee}{\end{equation}}
\newcommand{\bal}{\begin{aligned}}
\newcommand{\eal}{\end{aligned}}
\newcommand{\bea}{\begin{eqnarray}}
\newcommand{\eea}{\end{eqnarray}}
\newcommand{\ben}{\begin{enumerate}}
\newcommand{\een}{\end{enumerate}}
\newcommand{\bit}{\begin{itemize}}
\newcommand{\eit}{\end{itemize}}
\newcommand{\bde}{\begin{widetext}}
\newcommand{\ede}{\end{widetext}}

\renewcommand{\(}{\left(}
\renewcommand{\)}{\right)}
\renewcommand{\[}{\left[}
\renewcommand{\]}{\right]}

\newcommand{\mathsym}[1]{{}}

\definecolor{ao}{rgb}{0.0, 0.5, 0.0}
\definecolor{bostonuniversityred}{rgb}{0.8, 0.0, 0.0}

\DeclareUnicodeCharacter{2212}{-}
\pdfoutput=1

\begin{document}
\begin{titlepage}
\begin{flushright}
    TUM-HEP 1351/21\\
    PI/UAN-2021-692FT
\end{flushright}

\begin{center}
\vspace*{1cm}

{\Large\bf
Gauged Inverse Seesaw from Dark Matter
}

\vspace*{0.8cm}

{\bf Asmaa~Abada$^{a\!}$\footnote[1]{asmaa.abada@ijclab.in2p3.fr}, Nicol\'{a}s Bernal$^{b}$\footnote[2]{nicolas.bernal@uan.edu.co}, Antonio E. C\'{a}rcamo Hern\'{a}ndez$^{c,d,e}$\footnote[3]{antonio.carcamo@usm.cl},\\ Xabier Marcano$^{f}$\footnote[4]{xabier.marcano@tum.de} and Gioacchino Piazza$^{a}$\footnote[5]{gioacchino.piazza@ijclab.in2p3.fr}}

\vspace*{.5cm}
$^{a}$P\^ole Th\'eorie, Laboratoire de Physique des 2 Infinis Irène Joliot Curie (UMR 9012)\\
CNRS/IN2P3,
15 Rue Georges Clemenceau, 91400 Orsay, France

$^{b}$Centro de Investigaciones, Universidad Antonio Nariño\\ Carrera 3 Este \# 47A-15, Bogotá, Colombia

$^{{c}}$Departamento de F\'{\i}sica, Universidad T\'{e}cnica Federico Santa Mar\'{\i}a\\ Casilla 110-V, Valpara\'{\i}so, Chile\\
$^{{d}}$Centro Cient\'{\i}fico-Tecnol\'{o}gico de Valpara\'{\i}so, Casilla 110-V, Valpara\'{\i}so, Chile\\
$^{{e}}$Millennium Institute for Subatomic Physics at the High-Energy Frontier, SAPHIR, Chile

$^{{f}}$Physik-Department, Technische Universit\"at M\"unchen\\ James-Franck-Stra\ss e, 85748 Garching, Germany

\vspace*{1cm}
\begin{abstract}
\noindent
We propose an economical model addressing the generation of the Inverse Seesaw  mechanism  from the spontaneous breaking of a  local $U(1)_{B-L}$, with the Majorana masses of the sterile neutrinos radiatively generated from the dark sector. 
The field content of the Standard Model is extended by neutral scalars and fermionic singlets, and the gauge group is extended with a $U(1)_{B-L}$ and a discrete $\mathbb{Z}_4$ symmetries.
Besides dynamically generating the Inverse Seesaw and thus small masses to the active neutrinos, our model offers two possible dark matter candidates, one scalar and one fermionic, stable thanks to a remnant $\mathbb{Z}_2$ symmetry. Our model complies with bounds and constraints form dark matter direct detection, invisible Higgs decays and $Z'$ collider searches for masses of the dark sector at the TeV scale.

\vspace{2cm}

\end{abstract}
\end{center}
\end{titlepage}

\section{Introduction}
The origin of neutrino masses is one of the big open questions in Particle Physics and, consequently, a plethora of explanations have been proposed, see {\it e.g.} Ref.~\cite{Cai:2017jrq}. 
The type-I seesaw mechanism~\cite{Minkowski:1977sc, Yanagida:1979as, Glashow:1979nm, Mohapatra:1979ia, GellMann:1980vs, Schechter:1980gr, Schechter:1981cv} is maybe the simplest and most popular one, which requires the extension of the Standard Model (SM) particle field content with new neutral leptons, known as right-handed (RH) or sterile neutrinos. 
Although the most minimal seesaw realization capable of accounting for neutrino oscillation data~\cite{Esteban:2020cvm, deSalas:2020pgw} requires two RH neutrinos, the case of three RH neutrinos is particularly interesting, as it restores the balance between the number of quarks and leptons, canceling the gauge anomalies of $U(1)_{B-L}$. 
In this case, $U(1)_{B-L}$ can be promoted from being an accidental global symmetry of the SM to a local gauge group. 
This kind of models have been extensively studied in the literature, see {\it e.g.} Refs.~\cite{Davidson:1987mh, Iso:2009ss, Iso:2009nw, Datta:2013mta, Das:2015nwk, Das:2016zue, Nomura:2017vzp, Chauhan:2018uuy, Das:2019pua, Abdallah:2011ew}.

In order to accommodate sub-eV neutrino masses~\cite{Aker:2021gma}, the canonical type-I seesaw model requires either a high lepton number violating (LNV) scale, or tiny Yukawa couplings if realized at low scale. 
Either way, the phenomenology of this model is extremely suppressed, which makes it difficult to probe experimentally.
An alternative to have a low-scale realization with large Yukawa couplings is to consider additional sterile fermions and an approximate  symmetry. 
This is the case of the  the inverse seesaw (ISS)~\cite{Wyler:1982dd, Mohapatra:1986bd, GonzalezGarcia:1988rw} or the linear seesaw (LSS)~\cite{Barr:2003nn, Malinsky:2005bi} realizations, where the $B-L$ global symmetry is used to protect active neutrino masses, linking their smallness to small parameters that quantify the  breaking of the $B-L$ symmetry.  
Here, the neutral leptons are introduced in pairs of opposite $B-L$, so they cannot cancel the anomalies of $U(1)_{B-L}$.
Therefore, gauging low-scale seesaw models requires to extend the particle content with new exotic fermions, as it was done for instance in Refs.~\cite{Okada:2012np,Basso:2012ti, Kajiyama:2012xg, Ma:2014qra, DeRomeri:2017oxa, Geng:2017foe, Nomura:2017kih, Mondal:2021vou,Fernandez-Martinez:2021ypo}. 

Additionally, to generate tree-level Majorana masses for the active neutrinos,  low-scale seesaw models introduce a new scale that breaks explicitly lepton number symmetry either with a Majorana mass term (as in the ISS) or via a Yukawa interaction for the new fermion singlets (as done in the LSS). In these scenarios, LNV parameters are {\it ad-hoc} and are assumed  to be small, an hypothesis that, despite being technically natural and thus stable under radiative corrections, lacks of a more fundamental motivation.
Possible explanations for the origin of these parameters, in particular the small Majorana mass term in the ISS construction, have been proposed in several models~\cite{Ma:2009gu,Bazzocchi:2010dt, Law:2012mj, Fraser:2014yha, Ahriche:2016acx, CarcamoHernandez:2013krw, CarcamoHernandez:2017owh, CarcamoHernandez:2018hst, CarcamoHernandez:2018iel, Bertuzzo:2018ftf, Rojas:2019llr, CarcamoHernandez:2019eme, CarcamoHernandez:2019pmy, CarcamoHernandez:2019vih, CarcamoHernandez:2019lhv, Hernandez:2021uxx, CarcamoHernandez:2021iat, Nomura:2021adf, Hernandez:2021mxo}. 

In this work we explore the possibility of gauging the ISS model by adding only sterile neutrinos in the fermionic sector and, at the same time, providing a dynamical one-loop generation of the Majorana mass $\mu$ for the sterile neutrinos, which justifies its smallness. 
Furthermore, this model provides viable dark matter (DM) candidates either from the sterile neutrino sector or from the extended scalar one, which are actually the particles behind the generation of the $\mu$ mass.
As a consequence, the neutrino and dark sectors are connected in this setup. 

Our model considers an extension of the SM by a local $U(1)_{B-L}$ symmetry and an additional $\mathbb{Z}_4$ symmetry, which protects active neutrinos from obtaining tree-level masses and ensures the stability of the DM candidates.
The scalar sector is extended with three scalar singlets, $\sigma$ that breaks $U(1)_{B-L}$, $\chi$ that breaks $\mathbb{Z}_4$ to a conserved $\mathbb{Z}_2$ symmetry, and an extra inert one $\zeta$. 
We will assume that the breakings occur at the TeV scale in order to be consistent with LHC bounds on the new heavy vector boson~\cite{Aad:2019fac,  Sirunyan:2021khd}.
The fermionic sector contains new singlets: three RH neutrinos $N_R$ that are in charge of canceling the gauge anomalies of $U(1)_{B-L}$, and two additional pairs of sterile neutrinos $(\nu_R,\, \nu_S)$ with opposite $B-L$ numbers, so that the low-scale seesaw is realized without spoiling the anomaly cancellation. 
Due to the $\mathbb{Z}_4$ symmetry, the $N_R$ fields do not mix with the active neutrinos and the $\nu_{R,S}$ neutral leptons are not allowed to acquire Majorana mass terms, consequently the active neutrinos remain massless at the tree level, even after the spontaneous symmetry breaking of $U(1)_{B-L} \otimes \mathbb{Z}_4$. 
Nevertheless, our setup generates Majorana masses  for the $\nu_S$ fields radiatively at the one-loop level, providing the origin of its smallness, and triggering the ISS mechanism. 
Furthermore, we will see that this loop contribution is proportional to small parameters protected by an accidental symmetry, and thus they could additionally suppress the $\mu$ term.

Moreover, the model provides two different DM candidates, whose stability is ensured by the residual $\mathbb{Z}_2$ symmetry. 
Depending on the mass hierarchy, we could have a fermionic DM candidate from the lightest of the three $N_R$, or a scalar candidate from the lightest component of the inert scalar $\zeta$ (real $\zeta_R$ or imaginary $\zeta_I$ component). 
Interestingly, the same parameter needed to dynamically generate the $\mu$ term of the ISS  will break the mass degeneracy of $\zeta_R$ and $\zeta_I$, connecting thus  both the DM and neutrino sectors. It is worth stressing that the loop generating the $\mu$ term involves the new DM candidates, making our mechanism similar to the scotogenic model for neutrino masses~\cite{Ma:2006km}, {\it i.e.} our model provides a scotogenic origin of the ISS $\mu$ term. 
We find that in order to comply with neutrino data, DM direct detection searches, invisible Higgs decays and $Z'$ collider searches, the masses of the DM candidates are larger than few TeV. 
Furthermore, in the fermionic DM scenario, the $U(1)_{B-L}$ gauge coupling needs to be above $0.7$ and the $Z'$ gauge boson heavier than about 6-20~TeV.

The paper is organized as follows: in Section~\ref{Sec:model} we introduce the model, providing a detailed description of the scalar sector.
The fermionic sector is discussed in Section~\ref{Sec:Neutrino}, with details on the active and sterile neutrino mass generation.
The viability of both scalar and fermionic DM candidates is discussed in Section~\ref{Sec:DM}. Details on the anomaly cancellations, stability and unitarity conditions are collected  in the Appendices. We summarize our findings in Section~\ref{Sec:conclusions}.

\section{The model}
\label{Sec:model}

We propose an extension of the SM where the gauge symmetry is extended by the inclusion of spontaneously broken $U(1)_{B-L}$ gauge and $\mathbb{Z}_{4}$ discrete symmetries, and the particle content is enlarged with new scalar and fermionic singlets. 
The particle spectrum and their charge assignments are shown in Table~\ref{Themodel}. 
\begin{table}[t!]
\renewcommand{\arraystretch}{1.3}
\centering%
\begin{tabular}{|c||c|c|c|c|c|c|c|c|c|c|c|c|}
\hline
Field & $q_{iL}$ & $u_{iR}$ & $d_{iR}$ & $\ell_{iL}$ & $\ell_{iR}$ & $\nu _{R_k}$ & 
$\nu_{S_k}$ & $N _{R_i}$ & $\phi $ & $\zeta $ & $\sigma $ & $\chi $ \\ \hline\hline
$SU(3)_{C}$ & $\mathbf{3}$ & $\mathbf{3}$ & $\mathbf{3}$ & 
$\mathbf{1}$ & $\mathbf{1}$ & $\mathbf{1}$ & $\mathbf{1}$ & $\mathbf{1}$ & $\mathbf{1}$
& $\mathbf{1}$ & $\mathbf{1}$ & $\mathbf{1}$ \\ 
\hline
$SU(2)_{L}$ & $\mathbf{2}$ & $\mathbf{1}$ & $\mathbf{1}$ & 
$\mathbf{2}$ & $\mathbf{1}$ & $\mathbf{1}$ & $\mathbf{1}$ & $\mathbf{1}$ & $
\mathbf{2}$ & $\mathbf{1}$ & $\mathbf{1}$ & $\mathbf{1}$ \\ 
\hline
$U(1)_{Y}$ & $\frac{1}{6}$ & $\frac{2}{3}$ & $-\frac{1}{3}$ & 
$-\frac{1}{2}$ & $-1$ & $0$ & $0$ & $0$ & $\frac{1}{2}$ & $0$ & $0$ & $0$ \\ 
\hline
$U(1)_{B-L}$ & $\frac{1}{3}$ & $\frac{1}{3}$ & $\frac{1}{3}$ & $
-1 $ & $-1$ & $-1$ & $1$ & $-1$ & $0$ & $0$ & $-2$ & $0$ \\ 
\hline
$\mathbb{Z}_{4}$ & $1$ & $1$ & $1$ & $1$ & $1$ & $1$ & $-1$ & $i$ & $1$ & $-i$ & $-1$
& $-1$ \\ \hline
\end{tabular}%
\caption{Particle charge assignments under the $SU(3)_{C} \otimes
SU(2)_{L} \otimes U(1)_{Y} \otimes U(1)_{B-L} \otimes \mathbb{Z}_{4}$ symmetry.
Starting with the SM Higgs boson doublet $\phi$, the scalar sector is collected in the last four columns,  preceded by three columns for the new sterile fermions, while de SM fermions are collected on the four first ones. 
More details are provided in the main text.
The indices run as follows: $i=1,2,3$ and $k=1,2$. 
}
\label{Themodel}
\end{table}

The scalar sector contains three new electrically neutral scalar gauge singlets $\sigma$, $\chi $ and $\zeta$.
The spontaneous symmetry breaking (SSB) of the $U(1)_{B-L}$ gauge symmetry is triggered by the vacuum expectation value (VEV) of $\sigma$, which we will take above the TeV scale in order to comply with collider bounds on the $Z'$ gauge boson~\cite{Aad:2019fac,  Sirunyan:2021khd}.
On the other hand, the VEV of $\chi$ breaks the $\mathbb{Z}_{4}$ symmetry down to a preserved $\mathbb{Z}_{2}$ one, which will ensure the stability of the DM particles. 
Finally, the scalar singlet $\zeta$ does not acquire a VEV, providing thus a scalar DM candidate with the lightest of its CP-even and -odd components.
As we will see, $\zeta$ plays a key role in implementing the active neutrino mass generation mechanism, inducing the ISS $\mu$ term at the one-loop level. 

The fermionic sector is extended with new sterile fermions $\nu_R$, $\nu_S$ and $N_R$, all of them singlets under the SM group but with different $U(1)_{B-L}\otimes\mathbb{Z}_4$ charges, thus playing different roles in the model. 
The $N_R$ fields have the same $B-L$ charge as the SM leptons, allowing to cancel the gauge anomalies of  $U(1)_{B-L}$. 
Indeed, this requirement imposes the number of $N_R$ to be three, one per generation.\footnote{Notice that our choice for the anomaly cancellation is not the only possible one. 
Indeed, the general case would require that $n_{N_R}+n_{\nu_R}-n_{\nu_S}=3$, with $n_i$ the number of each species. 
Our particular choice of $n_{N_R}=3$ and $n_{\nu_R}=n_{\nu_S}$ is to accommodate a low-scale seesaw realization in this gauged $B-L$ framework. }
Due to the $\mathbb Z_4$ symmetry, they do not mix to the active neutrinos, preventing them from generating tree-level neutrino masses and providing a fermionic DM candidate, the lightest of the three states $N_R$. On the other hand, the $\nu_R$ and $\nu_S$ fields have opposite $B-L$ charges and do not contribute to the anomaly cancellation of $B-L$ as long as they are introduced in pairs.
Interestingly, this same condition allows us to implement an ISS mechanism for the neutrino mass generation.
Notice that in principle we do not have any constrain on the number of $(\nu_R, \nu_S)$ pairs in the model, nevertheless we will consider only two pairs, the minimal ISS scenario able to accommodate neutrino oscillation data~\cite{Abada:2014vea}.

Regarding the neutral fermion masses, given the charge assignments displayed in Table~\ref{Themodel}, they are all dynamically generated by the SSB of $U(1)_{B-L}\otimes\mathbb{Z}_4$.
The scalar $\sigma$ provides Majorana masses for the $N_R$ fields and the $\chi$ generates Dirac-like masses for the $(\nu_R,\nu_S)$ pairs. 
Notice that the $\mathbb{Z}_4$ symmetry prevents the $\nu_R$ and $\nu_S$ fields from acquiring Majorana masses at tree level, and thus keeps the active neutrinos massless.
Nevertheless, as it is discussed later, the $N_R$ fields together with the real and imaginary parts of scalar singlet $\zeta $ generates a Majorana mass for $\nu_S$ at the one-loop level, triggering an ISS mechanism responsible for the light (active) neutrino masses.

\subsection*{The scalar potential}
The most general scalar potential invariant under the symmetries of the model reads
\be
\bal
V&= m ^2 \phi^\dagger \phi + m_\sigma^2 \sigma^* \sigma + \frac{1}{2}m_\chi^2 \chi^2 + m_\zeta^2\zeta^* \zeta \\
&+ \lambda \(\phi^\dagger \phi\)^2+\lambda_\sigma\(\sigma^* \sigma\)^2 + \lambda_\zeta\(\zeta^* \zeta\)^2+  \frac{1}{4}\lambda_\chi\chi^4 \\
&+\phi^\dagger \phi \[ \lambda_{\phi\sigma}\sigma^* \sigma+\lambda_{\phi\zeta}\zeta^* \zeta + \lambda_{\phi\chi}\chi^2\] + \zeta^* \zeta \[ \lambda_{\zeta \sigma} \sigma^* \sigma + \lambda_{\zeta \chi}\chi^2\] + \lambda_{\sigma \chi}  \sigma^* \sigma \chi^2\\
&+ \eta_\mu \zeta \zeta \chi  + \lambda'_\zeta\(\zeta \zeta\)^2 + h.c.
\eal
\label{Eq:potential}
\ee
Notice that the $\chi$ field, given its charge assignment, can be taken to be real.
Since we are interested in the case where the fields $\phi, \sigma$ and $\chi$ acquire  VEVs ($v$, $v_\sigma$ and $v_\chi$, respectively), but $\zeta$ does not, we consider in Eq.~(\ref{Eq:potential}) $m ^2$, $m_\sigma^2$ and $m_\chi^2$ to be negative, whereas $m_\zeta^2$ is taken  positive.
We focus on the CP-conserving case where the dimensionful trilinear $\eta_\mu$ and the quartic couplings are real. Additional constraints on the parameter space come from stability and unitarity conditions, which are discussed and summarized in Appendix~\ref{sec:stability-unitarity}. 
The scalar fields are given by,
\be
\bal
& \phi =\begin{pmatrix} \phi^+\\
\frac{1}{\sqrt{2}}( v+ h + i \phi_Z )
\end{pmatrix}, \quad 
&&\sigma=\frac{1}{\sqrt{2}} \(v_\sigma + \Tilde{\sigma} + i \sigma_{Z'}\),\\
&\chi = v_\chi +\Tilde{\chi}\ , \quad 
&&\zeta = \frac{1}{\sqrt{2}} \(\zeta_R + i \zeta_I\),
\eal
\ee
where the VEVs $v$, $v_\sigma$ and $v_\chi$ break the electroweak symmetry, $U(1)_{B-L}$ and $\mathbb{Z}_4$, respectively.
Here, $h, \Tilde{\sigma}$ and $\Tilde\chi$ are the three physical scalars remnant from each SSB, while $\phi^\pm$, $\phi_Z$ and $\sigma_{Z'}$ are the Goldstone bosons associated to the  longitudinal components of the gauge fields $W^\pm$, $Z$ and $Z'$, respectively.
Minimizing the scalar potential, we find that the VEVs of the scalar fields are solutions of the following equations 
\begin{align} 
&2\lambda  v^2+2\lambda_{\phi\chi} v_\chi^2+\lambda_{\phi\sigma}v_\sigma^2+2 m^2=0\, \label{eq:v1} ,\\
&2\lambda _\sigma v_\sigma^2+2\lambda_{\sigma \chi}v_\chi^2+\lambda_{\phi\sigma}v^2+2 m_\sigma^2=0\, \label{eq:v2},\\
& \lambda _\chi v_\chi^2+\lambda_{\sigma \chi}v_\sigma^2+\lambda_{\phi\chi} v^2+ m_\chi^2=0\, . \label{eq:v3}
\end{align}
After SSB, the $\mathbb Z_4$ is broken down to a residual $\mathbb Z_2$ under which $h, \Tilde\sigma, \Tilde\chi$ are even, and $\zeta_R, \zeta_I$ are odd, thus decoupling the two sectors. 
We can write the mass matrix for the $\mathbb Z_2$-even states as 
\be
\bal
-\mathcal{L}_\text{Mass}^{+} = \frac{1}{2}\left( h, \Tilde{\sigma}, \Tilde{\chi} \right) 
\left(
\begin{array}{ccc}
    2 \lambda  v^2 &  \lambda_{\phi \sigma} v v_\sigma & 2 \lambda_{\phi \chi} v_\chi v \\
    \lambda_{\phi \sigma} v v_\sigma & 2 \lambda _\sigma v_\sigma^2 & 2\lambda_{\sigma \chi} v_\chi v_\sigma \\
    2\lambda_{\phi \chi} v_\chi v & 2 \lambda_{\sigma \chi} v_\chi v_\sigma & 2\lambda _\chi v_\chi^2 \\
\end{array}
\right)
\left( 
\begin{array}{c}
    h\\
    \Tilde{\sigma}\\
    \Tilde{\chi}%
\end{array}%
\right).
\eal
\ee

The quartic couplings $\lambda_{\phi\sigma}$ and $\lambda_{\phi\chi}$ control the size of the deviations between $ h$ and the SM Higgs boson,  which are strongly suppressed from LHC Higgs data~\cite{Zyla:2020zbs}.
Consequently, we will focus on the decoupling scenario $\lambda_{\phi\sigma} = \lambda_{\phi\chi}=0$, in which $ h$ behaves as the SM Higgs. 
Then, one can analytically find the masses for the eigenstates $\Tilde{\sigma}'$ and $\Tilde{\chi}'$, defined as
\be
\bal
\left( 
\begin{array}{c}
\Tilde{\sigma}'\\
\Tilde{\chi}'%
\end{array}%
\right) = \left(
\begin{array}{cc}
\cos{\theta} & -\sin{\theta}\\
\sin{\theta} &\cos{\theta} \\
\end{array}
\right)
\left( 
\begin{array}{c}
\Tilde{\sigma}\\
\Tilde{\chi}%
\end{array}%
\right)\,,
\eal
\ee
with 
\begin{align}
    m_{\Tilde{\sigma}'}^2 &= \lambda _\chi v_\chi^2+\lambda _\sigma v_\sigma^2+ \sqrt{\left(\lambda_\sigma v_\sigma^2 -\lambda_\chi v_\chi^2\right)^2 + 4 \lambda_{\sigma\chi}^2 v_\sigma^2 v_\chi^2}\,,\\
    m_{\Tilde{\chi}'}^2 &=\lambda _\chi v_\chi^2+\lambda _\sigma v_\sigma^2-
    \sqrt{\left(\lambda_\sigma v_\sigma^2 -\lambda_\chi v_\chi^2\right)^2 + 4 \lambda_{\sigma\chi}^2 v_\sigma^2 v_\chi^2}\,,\\
    \tan{ 2 \theta} &= \frac{2 \lambda_{\sigma\chi } v_\chi v_\sigma}{\left(\lambda _\sigma v_\sigma^2-\lambda _\chi v_\chi^2\right)}\,.
\end{align}
On the other hand, the masses of the real and imaginary part of the $\zeta$ field are given by
\begin{align}
    &m_{\zeta_R}^2=\frac{1}{2}\left(2 \lambda_{\zeta \chi} v_\chi^2+\lambda_{\zeta \sigma} v_\sigma^2+\lambda_{\phi \zeta} v^2+2 m_\zeta^2 +2 v_\chi \eta_\mu \right),\\
    &m_{\zeta_I}^2=\frac{1}{2}\left(2 \lambda_{\zeta \chi} v_\chi^2+\lambda_{\zeta \sigma} v_\sigma^2+\lambda_{\phi \zeta} v^2+2 m_\zeta^2 -2 v_\chi \eta_\mu \right),
\end{align}
and their  squared mass difference is proportional to the dimensionful  parameter $\eta_\mu$, 
\begin{equation}
    m_{\zeta_R}^2-m_{\zeta_I}^2= 2 v_\chi \eta_\mu\, .
    \label{Eq:splitting}
\end{equation}
Therefore, depending on the sign of $\eta_\mu$, either of $\zeta_R$ or $\zeta_I$ could be a viable DM candidate, whose stability is ensured as they are odd under the residual $\mathbb{Z}_2$ symmetry.
As we will see in the next section, the same parameter $\eta_\mu$ is the key ingredient behind  the ISS mechanism, as it will control the dynamical generation of the one-loop Majorana mass for the sterile neutrinos.

\section{Active and Sterile neutrino mass generation}\label{Sec:Neutrino}

With the symmetries and particle content of our model, the neutrino Yukawa terms in the Lagrangian are given by:
\begin{equation}
-\mathcal{L}_{Y}^{\left( \nu \right)}=
y_\phi^{ik}\,\bar\ell_{iL}\widetilde{\phi }\nu _{R_k}
+y_\chi^{kr}\,\bar\nu _{R_k} \nu_{S_r}^{C}\chi 
+y_\zeta^{ki}\,\bar{\nu}_{S_k} N _{R_i}^{C}\zeta
+y_\sigma^{ij}\,\bar N_{R_i}N_{R_j}^{C}\sigma +h.c.
\end{equation}
Here,  the superscript $C$ stands for charge conjugation $\Psi^C= C \overline{\Psi}^T$, $\widetilde{\phi }=i \tau_2 \phi^*$, and summation over repeated indices must be understood with $i,j=1,2,3$ and $k,r=1,2$.
After the SSB of the local $SU(2)_{L}\otimes U(1)_{Y}\otimes U(1) _{B-L} \otimes \mathbb Z_4$ symmetry we obtain a Majorana mass matrix for the neutrinos, which in the basis $(\nu_L, \nu_R^C, \nu_S^C, N_R^C)^T$ has the following structure,
\begin{equation}
M_{\nu} \sim \left(\begin{array}{ccc;{3pt/3pt}c}
0 & m_{D} & 0 & 0\\ 
m_{D}^{T} & 0& M & 0\\ 
0 & M^{T} & \mu & 0\\ 
\hdashline[3pt/3pt]
 0 & 0 & 0 &m_{N}
\end{array}\right)\,. \label{Eq:mu-entries}
\end{equation}
Notice that it is separated in two sectors: a low-scale seesaw one and a decoupled Majorana mass matrix $m_N$.
The dimensions of the submatrices are $3\times 2$ for $m_D$, $2\times2$ for $M$ and $\mu$, and $3\times3$ for $m_{N}$. 
At tree level, they are given by 
\begin{equation}
m_D=y_\phi \frac v{\sqrt2}\,, \quad
M= y_\chi v_\chi\,,\quad
m_{N}= y_{\sigma} \frac{v_\sigma}{\sqrt2}\quad
{\rm and}\quad
\mu=0\,.
\end{equation} 
The structure of the mass matrix in Eq.~\eqref{Eq:mu-entries} is a consequence of the $\mathbb{Z}_4$ charge assignment, which also forbids a tree-level mass for $\nu_S$ even after the SSB. 
In this case of $\mu=0$, the active neutrinos remain massless, as the $(\nu_R, \nu_S)$ neutrinos become degenerate, forming a Dirac neutrinos of mass $M$, and their contribution to the light active neutrino masses vanishes. 
On the other hand, the Majorana neutrinos $N_R$ are decoupled from the active neutrinos, again due to the $\mathbb{Z}_4$ symmetry, they  therefore do not contribute to the active neutrino masses at tree level.
They are however responsible for generating the Majorana $\mu$  term at one-loop level. 

\begin{figure}[t!]
\centering
\includegraphics[width=0.6\textwidth]{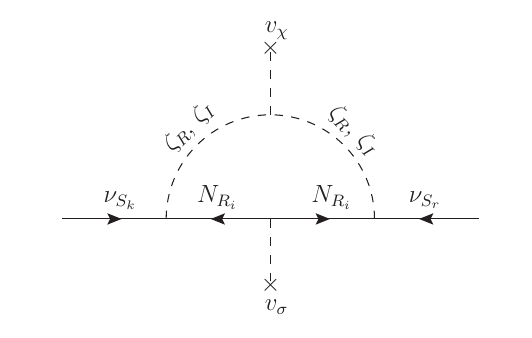}
\caption{Feynman diagram for the generation of the Majorana neutrino mass $\mu$ via a loop involving the new particles from the dark sector, $N_R$ and $\zeta$.
The indices run as follows: $i=1,2,3$ and $k,r=1,2$.}\label{Loopdiagramsmu}
\end{figure}

We show in Fig.~\ref{Loopdiagramsmu} the relevant diagram generating the Majorana mass term for the $\nu_S$ sterile neutrinos, which involves a loop with the $N_R$ fields and the inert scalar $\zeta$.   
Notice that both real and imaginary components of the scalar singlet $\zeta$ participate in the diagram, coupled to the scalar singlet $\chi$ via the trilinear  coupling  $\eta_\mu$ defined in Eq.~\eqref{Eq:potential}, which is at the same time responsible for generating the mass splitting between $\zeta_R$ and $\zeta_I$ in Eq.~\eqref{Eq:splitting}.
From this diagram, we obtain the  one-loop contribution given by
\begin{equation}
\mu_{\rm 1-loop}^{kr} =\frac{y_{\zeta}^{ki} y_{\zeta}^{ri} m_{N_i}}{16\pi ^{2}}
\,\left\{ \frac{m_{\zeta _{R}}^{2}}{m_{\zeta _{R}}^{2}- m_{N_i }^{2}}\log \left( \frac{m_{\zeta _{R}}^{2}}{m_{N_i }^{2}}\right) 
-\frac{m_{\zeta _{I}}^{2}}{m_{\zeta_{I}}^{2}-m_{N_i }^{2}}\log \left( \frac{m_{\zeta _{I}}^{2}}{m_{N_i }^{2}}\right)
\right\}\,,
\end{equation}%
where we have assumed, without lost of generality, that the submatrix $m_{N}$ is diagonal. 
This non-zero $\mu$ matrix triggers the ISS mechanism in the $(\nu_L, \nu_R^C, \nu_S^C)$ sector, generating masses for the active neutrinos of the order%
\footnote{Here we discuss for the sake of simplicity the order of magnitude in the case of one generation. The generalization to more families is straightforward.}
$m_\nu\sim \mu\, m_D^2/M^2$ and mass splittings for the pairs of sterile neutrinos of $\mathcal O(\mu)$,  thus implying that the sterile neutrinos form pseudo-Dirac pairs.

As usual, a successful ISS realization requires a light $\mu$ scale, in particular $\mu\ll m_D, M$, which  occurs naturally in our model. 
In order to explore this feature in detail, let us start assuming a small mass splitting between the real and imaginary components of $\zeta$, {\it i.e.} 
\begin{equation}m_{\zeta_R}^2-m_{\zeta_I}^2 = 2 v_\chi \eta_\mu \ll 2 m_\zeta^2\equiv m_{\zeta_R}^2+m_{\zeta_I}^2\ .
\label{Eq:small-splitting}
\end{equation}
Then, we can write
\begin{equation}\label{Eq:mu1loopapprox}
\mu_{\rm 1-loop}^{kr} \simeq \frac{y_{\zeta}^{ki} y_{\zeta}^{ri}}{8\pi ^{2}}
\, v_\chi \eta_\mu\, m_{N_i}\,\frac{m_\zeta^2-m_{N_i}^2+m_{N_i}^2 \log m_{N_i}^2/m_\zeta^2}{(m_\zeta^2-m_{N_i}^2)^2}\,.
\end{equation}
From this equation, we see that the generated $\mu$ scale is suppressed by a loop factor, and it is proportional to the $y_\zeta$ and $\eta_\mu$ parameters. 
Interestingly, in the limit when these parameters are zero, together with the quartic coupling $\lambda'_\zeta$, the Lagrangian becomes invariant under a global $U(1)_\zeta$ symmetry under which only the $\zeta$ fields are charged. 
In this sense, it is technically natural to consider these three parameters, and in particular $y_\zeta$ and $\eta_\mu$, to be small, a fact that further suppresses the $\mu$ scale. 
Therefore, our model predicts a small scale for the ISS $\mu$ term, which is suppressed by a loop factor and three powers of technically natural small parameters. 

Furthermore, Eq.~\eqref{Eq:mu1loopapprox} exhibits the link between the light neutrino masses and the dark sector, and can be used to relate the relevant scales for both sectors. 
In order to do that, let us recall again that in the ISS mechanism the light active neutrino mass scale $m_\nu$ is of the order of
\begin{equation}
m_\nu \approx \mu\, \frac{m_D^2}{M^2}\approx \mu\, \big|U_{\nu N}\big|^2\,,
\end{equation}
with $U_{\nu N}$ the active sterile neutrino mixing. 
This mixing modifies the neutral and charged lepton currents,  leading thus to numerous constraints. 
Cosmological observations~\cite{Smirnov:2006bu, Kusenko:2009up, Hernandez:2014fha, Vincent:2014rja} 
put severe constraints on neutral leptons  with a mass below $\sim 200$~MeV.
Additional strong bounds for sterile neutrinos lighter that the mass of the  W gauge  boson are imposed from several laboratory searches, see for instance Refs.~\cite{Atre:2009rg, Abada:2017jjx, Bolton:2019pcu}.
In the following, we focus on the case where the sterile neutrinos realizing the ISS mechanism are heavier than the electroweak scale, which are mainly constrained by electroweak observables and charged lepton flavor transitions~\cite{Abada:2018nio}.
A global fit analysis to these observables~\cite{Fernandez-Martinez:2016lgt} imposes bounds on their mixings of the order $|U_{\nu N}|^2\lesssim 10^{-3}$, which implies that $\mu\gtrsim \mathcal O(1)$~keV in order to reproduce $\sim 1$~eV neutrino masses. 

On the other hand, Eq.~\eqref{Eq:mu1loopapprox} can be studied in two interesting limits, both related to the mass hierarchy between the lightest $N_{R}$ and $\zeta$ defining the DM candidate in our model.
In the case where $m_N\ll m_\zeta$, the model contains a fermionic DM candidate with mass $m_N$, and the $\mu$ scale of the ISS can be expressed, using Eq.~\eqref{Eq:small-splitting}, as
\begin{equation}
\mu\approx \frac{y_\zeta^2}{16\pi^2}\,\frac{\Delta m_\zeta^2}{m_\zeta^2}\, m_N 
\approx 6\,{\rm keV}~\bigg(\frac{y_\zeta}{10^{-2}}\bigg)^2\, 
\bigg(\frac{\Delta m_\zeta^2/m_\zeta^2}{10^{-2}}\bigg)\,
\bigg(\frac{m_N}{1\,{\rm TeV}}\bigg)\,. 
\end{equation}
In the opposite limit of $m_\zeta\ll m_N$, we have a scalar DM with mass $m_\zeta$, and 
\begin{equation}
\mu\approx \frac{y_\zeta^2}{16\pi^2}\,\frac{\Delta m_\zeta^2}{m_\zeta^2}\, \frac{m_\zeta^2}{m_N} 
\Big(\log\frac{m_N^2}{m_\zeta^2}-1\Big) 
\approx 1\,{\rm keV} \bigg(\frac{y_\zeta}{10^{-2}}\bigg)^2\, 
\bigg(\frac{\Delta m_\zeta^2/m_\zeta^2}{10^{-2}}\bigg)\,
\bigg(\frac{m_\zeta}{1\,{\rm TeV}}\bigg)^2
\bigg(\frac{10\,{\rm TeV}}{m_N}\bigg)\,,
\end{equation}
where we have neglected the logarithmic dependence in the last step. Notice that the case  $m_N\sim m_\zeta$ leads to the same $\mu$ term than in the case  $m_N\ll m_\zeta$, with just an additional factor of $1/2$.
From the above equations, we see that our model can easily generate the correct scale for the ISS $\mu$ term, with a DM candidate at the TeV scale, and relatively small $y_\zeta$ and $\Delta m_\zeta^2/m_\zeta^2$ (and thus $\eta_\mu$), which are protected by the accidental $U(1)_\zeta$ symmetry. 

In conclusion, our model connects the ISS $\mu$ term with the dark sector via the loop diagram in Fig.~\ref{Loopdiagramsmu}, in a similar fashion as the scotogenic model~\cite{Ma:2006km} does for neutrino mass generation. 
Notice however that in the original scotogenic model the dark sector generates directly small radiative masses for the active neutrinos, while in our model the dark sector generates only the Majorana masses for the sterile neutrinos $\nu_S$, which then induce masses for active neutrinos via the ISS mechanism. 
In this sense, our model provides a scotogenic origin for the ISS $\mu$ term.
A similar idea has been proposed, for instance, in Refs.~\cite{Ma:2009gu, Fraser:2014yha, Ahriche:2016acx, Rojas:2019llr}, with different particle content and symmetries than ours, and without linking it to the gauging of $U(1)_{B-L}$, as done in this work.

\section{Dark matter}
\label{Sec:DM}
The residual $\mathbb{Z}_2$ symmetry protects the lightest odd state of the dark sector, rendering it a viable candidate for DM.
Depending on the mass hierarchy, we could have scalar or fermionic DM candidates if $\zeta$ or $N_R$ are the lightest states, respectively.
The phenomenology of the WIMP paradigm of these two cases will be studied in the following.

\subsection{Scalar dark matter}
\begin{figure}[t!]
	\centering
    \includegraphics[width=16cm]{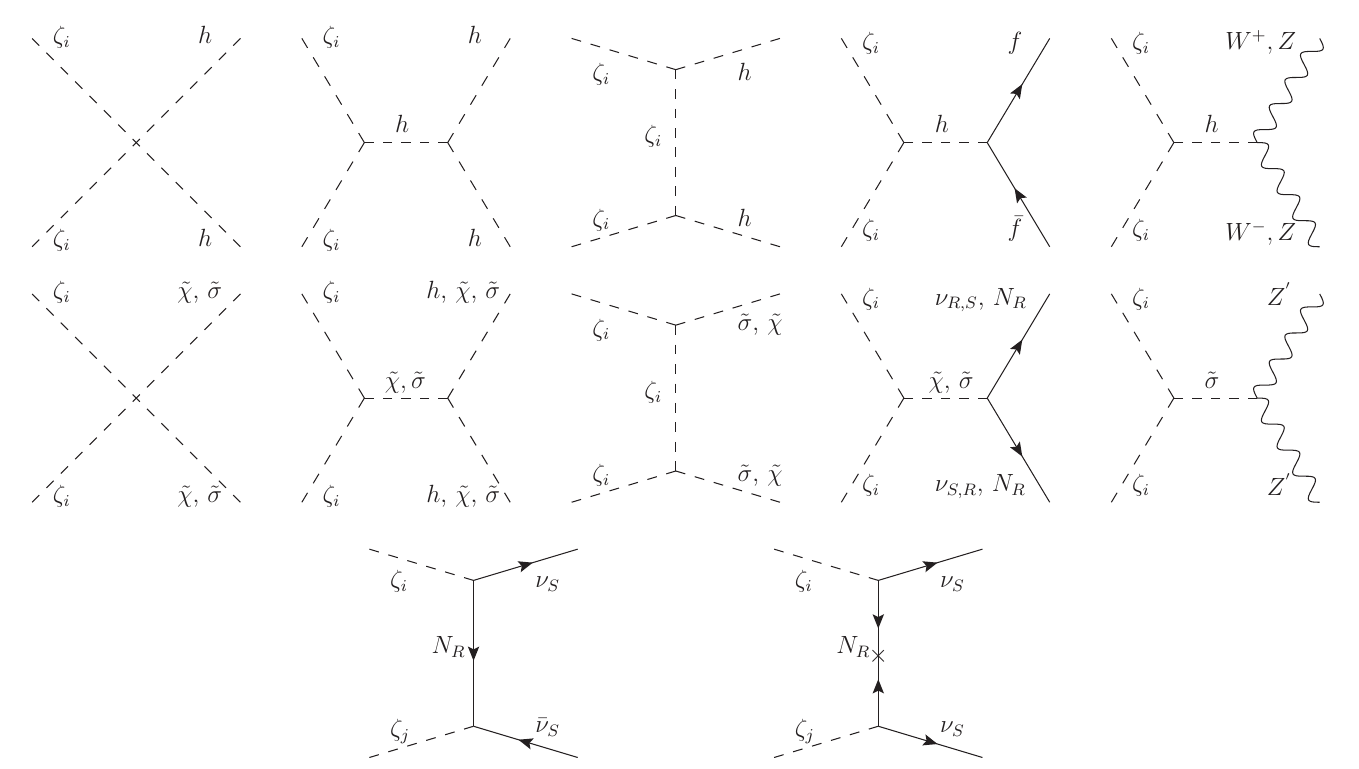}
    \caption{Main channels contributing to the production of {\it scalar} DM $\zeta_i$, with $i$ identifying the lightest state between the real and imaginary part of $\zeta$, and $j=R$, $I$. Crossed diagrams are not shown, although taken into account in the analysis. $f$ labels all the fermion fields coupled to the Higgs. 
    }
    \label{fig:DiagScalarDM}
\end{figure}
In this section, we consider the case where the DM candidate is the lightest component of $\zeta$, either its real or its imaginary part.
In the early Universe, it can be produced out of the SM thermal bath by the WIMP mechanism, via the different 2-to-2 scatterings presented in Fig.~\ref{fig:DiagScalarDM}.
DM can annihilate into SM Higgs bosons via the contact interaction, the $s$-channel exchange of a Higgs, or the $t$- and $u$-channel mediation of a $\zeta$.
It is also possible to have a scattering into SM charged fermions and vector bosons, mediated by the Higgs portal.
Additionally, the new scalars $\tilde \sigma$ and $\tilde \chi$ could be in the final state or mediate the annihilation, as shown in the second row of Fig.~\ref{fig:DiagScalarDM}.
Finally, DM can annihilate into neutrinos via the $t$-channel exchange of $N_R$ (last row).
This latter channel exists not only for the DM annihilation but also for the coannihilation $\zeta_R$-$\zeta_I$, which is typically relevant if the mass splitting, see Eq.~(\ref{Eq:splitting}), is smaller that $\sim 20\%$~\cite{Griest:1990kh}.
However, it is interesting to note that both the annihilation and the coannihilation into  active neutrinos depends on the active-sterile mixing angles (to the fourth power), which are suppressed at high temperature ({\it i.e.} at $T \gtrsim 200$~MeV) due to over dominating thermal masses of the active neutrinos~\cite{Dolgov:2000ew, Drewes:2016upu, Lello:2016rvl}, and therefore these channels are subdominant.

\begin{figure}
	\centering
	\includegraphics[width=0.6\textwidth]{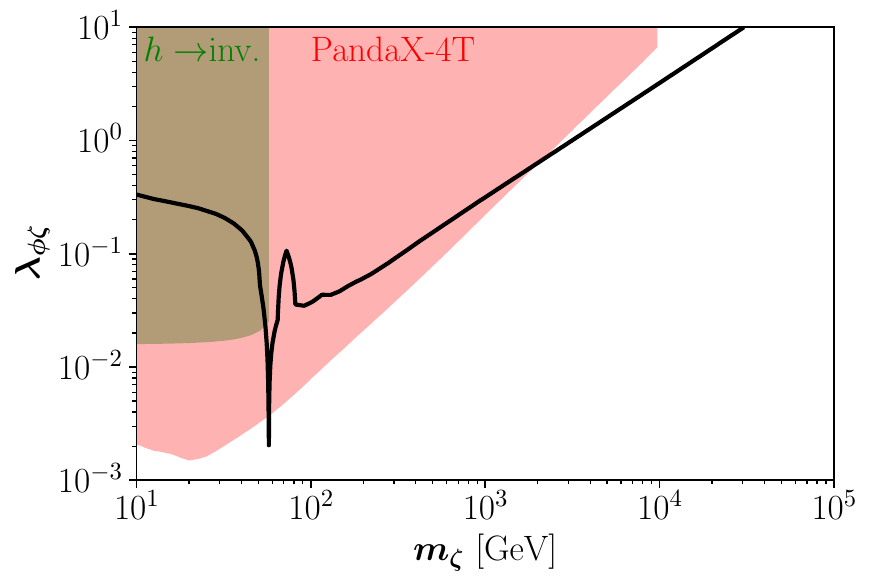}
    \caption{Scalar DM. Parameter space that reproduces the whole observed DM abundance (black thick line).
    The color-shaded areas are in tension with the invisible decay of the Higgs (brown) or PandaX-4T results (red).
    }
	\label{fig:ScalarDM}
\end{figure}

In the decoupling limit ($\lambda_{\phi\sigma} = \lambda_{\phi\chi} = 0$), where $h$ behaves as the SM Higgs, and if $\zeta$ is lighter than the other dark sector particles, the main annihilation channels correspond to the ones on the first row of Fig.~\ref{fig:DiagScalarDM}, and the present scenario reduces to the scalar singlet DM model~\cite{Silveira:1985rk, McDonald:1993ex, Burgess:2000yq}.
Figure~\ref{fig:ScalarDM} shows with a black line the values for the Higgs portal coupling $\lambda_{\phi\zeta}$ as a function of the DM mass $m_\zeta$ required to reproduce the whole observed DM abundance.
A full numerical computation  has been performed using MicrOMEGAs~\cite{Belanger:2006is, Belanger:2018ccd}.%
\footnote{For the numerical analysis we have assumed the real part of $\zeta$ to be  lighter than its imaginary part, however the results remain in the opposite case.}

This scenario is constrained by collider measurements~\cite{Barger:2007im, Djouadi:2011aa, Djouadi:2012zc, Damgaard:2013kva, No:2013wsa, Robens:2015gla, Han:2016gyy}.
In particular, strong bounds on the invisible decay of the Higgs apply for $m_\zeta \lesssim 60\ \text{GeV}$, as one can have $h \to \zeta_R \zeta_R$ and $h \to \zeta_I \zeta_I$.
The corresponding decay width is given by
\begin{equation}
    \Gamma_{h \to \zeta\zeta} = \frac{\lambda_{\phi\zeta}^2}{32\pi} \frac{v^2}{m_h} \left[\sqrt{1 - 4\frac{m_{\zeta_R}^2}{m_h^2}} + \sqrt{1 - 4\frac{m_{\zeta_I}^2}{m_h^2}}\right].
\end{equation}
A recent combination of searches with the ATLAS experiment found an upper bound for this branching ratio $\text{Br}(h\to\text{inv}) < 0.11$ at 95\%~CL~\cite{ATLAS:2020kdi}.
The corresponding excluded region is overlaid in brown  in Fig.~\ref{fig:ScalarDM}. 

The DM direct detection can further constrain the parameter space.
The corresponding spin-independent DM-nucleon scattering cross section is given by~\cite{He:2009yd, Baek:2014jga, Feng:2014vea, Han:2015hda, Athron:2018hpc}
\begin{equation}
    \sigma \simeq \frac{\lambda_{\phi\zeta}^2\, f_n}{4\pi} \frac{\mu_\zeta^2\, m_n^2}{m_h^4\, m_\zeta^2}\,,
\end{equation}
where $m_n$ is the nucleon mass, $\mu_\zeta \equiv m_\zeta\, m_n / (m_\zeta + m_n)$ is the DM-nucleon reduced mass, and $f_n \sim 0.3$ is the hadron matrix element.
In Fig.~\ref{fig:ScalarDM} the recent limits from the PandaX-4T experiment~\cite{Meng:2021mui} are shown in red.

Additionally, indirect detection with $\gamma$-rays and antimatter could further constrain this scenario~\cite{Yaguna:2008hd, Goudelis:2009zz, Profumo:2010kp, Cline:2013gha, Urbano:2014hda, Duerr:2015mva, Duerr:2015aka, Benito:2016kyp}, however the limits are comparable to the ones coming from direct detection and thus we do not show them in Fig.~\ref{fig:ScalarDM}.
The combination of the direct detection bound together with the invisible decay width excludes a large fraction of the parameter space, letting unconstrained the narrow mass window close to the Higgs funnel ($m_\zeta \sim m_h/2$), and the region above the TeV, {\it i.e.}, 2~TeV~$\lesssim m_\zeta \lesssim 30$~TeV.
Notice that larger masses would require non-perturbative $\lambda_{\phi\zeta}$ in order to not overclose the Universe.

Before closing this section, it is interesting to note that in the present scenario alternatives to the WIMP mechanism exists.
For example, DM could also have been produced non-thermally in the early Universe via the FIMP mechanism~\cite{McDonald:2001vt, Choi:2005vq, Hall:2009bx, Elahi:2014fsa, Bernal:2017kxu}.
In that case, much smaller Higgs portal couplings are required $\mathcal{O}(\lambda_{\phi\zeta}) \sim 10^{-11}$ with a broader mass range~\cite{Yaguna:2011qn, Campbell:2015fra, Kang:2015aqa}.
Additionally, if the Higgs portal coupling is suppressed and DM has sizable self-interactions due to a quartic coupling of order $\mathcal{O}(\lambda_\zeta) \sim 1$, thermalization in the dark sector could have a strong impact on the DM abundance~\cite{Bernal:2015xba, Heikinheimo:2017ofk, Bernal:2020gzm}.
Moreover, DM could have been Hawking radiated by primordial black holes~\cite{Bernal:2020bjf}.
Finally, this model has also been studied in the framework of non-standard cosmologies~\cite{Bernal:2018ins, Hardy:2018bph, Bernal:2018kcw, Allahverdi:2020bys}.

\subsection{Fermionic dark matter}
\begin{figure}[t!]
	\centering
    \includegraphics[width=16cm]{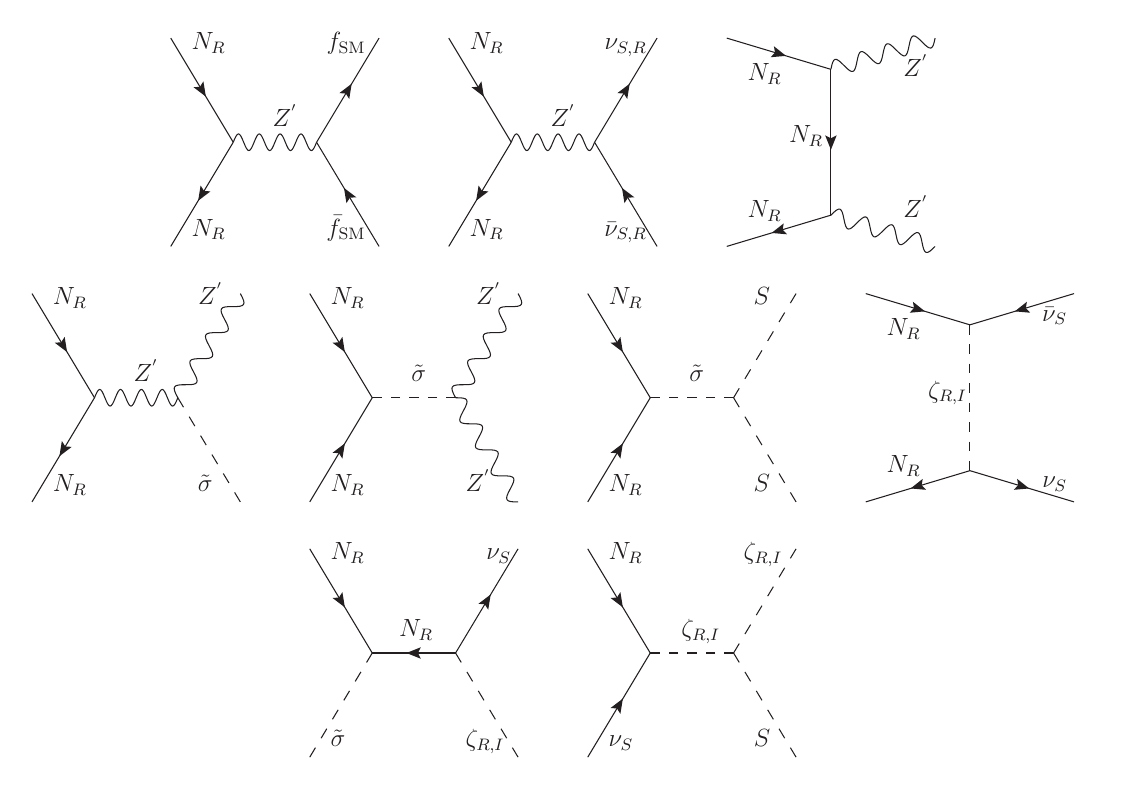}
    \caption{Main channels contributing to the production of {\it fermionic} DM. $f_{\text{SM}}$ indicates all the SM fermions coupled to the $Z'$ and $S=h$, $\tilde \sigma$, $\tilde \chi$, and $\zeta_{R,I}$.}
    \label{fig:DMfermion}
\end{figure}
In this case, the DM candidate is the lightest $N_{R_i}$ state.
In the early Universe, DM can  annihilate into $a)$ SM fermions and sterile neutrinos $\nu_R$ and $\nu_S$ mediated via the $s$-channel exchange of a $Z'$, together with the annihilation into a pair of $Z'$ (first row in Fig.~\ref{fig:DMfermion}), or $b)$ $Z'\, \sigma$, $Z'\, Z'$, a couple of scalars or neutral fermions (second row).
Finally, $c)$ DM can coannihilate with $\tilde \sigma$ or an active neutrino via the active-sterile mixing (last row).
As in the case of scalar DM, here some processes are typically subdominant because the coupling $\lambda_{\phi\sigma}$ has to be very small to avoid tension with Higgs physics, and because the active-sterile mixing angles are suppressed at high temperatures.

\begin{figure}
	\centering
	\includegraphics[width=0.7\textwidth]{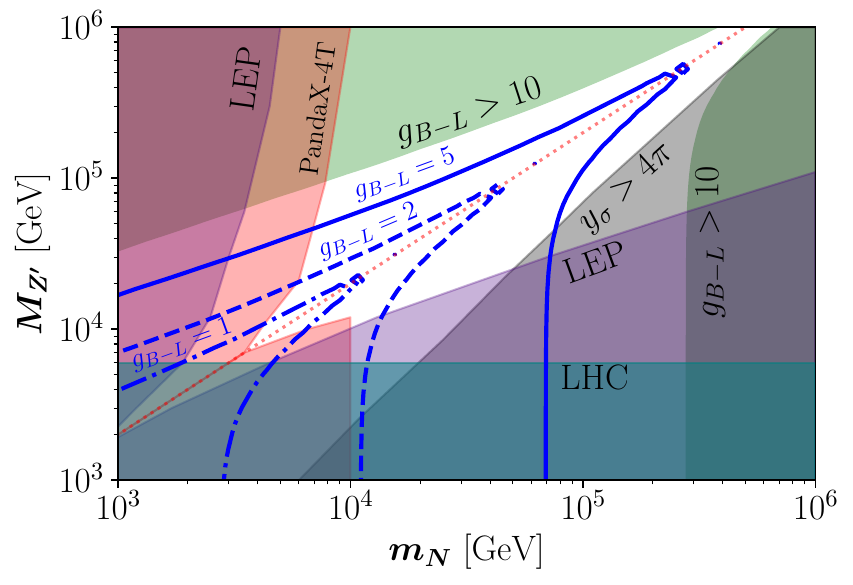}
    \caption{Fermionic DM. Contours for the coupling $g_{B-L}$ required in order to reproduce the whole observed DM abundance (thick blue lines).
    The color-shaded areas are in tension with LHC searches for $Z'$ gauge bosons (blue), LEP measurements of $e^+e^-\to\ell^+\ell^-$ (purple), direct DM searches from PandaX-4T (red), and the perturbative  bounds on $g_{B-L}$ (green) and $y_\sigma$ (gray).
    The red dotted line corresponds to $2\, m_N = M_{Z'}$.
    }
	\label{fig:gX}
\end{figure}

If the DM $N_R$ is the lightest state of the dark sector, most of the channels described in Fig.~\ref{fig:DMfermion} are kinematically closed, and therefore the annihilation into SM states via the exchange of a $Z'$ boson tends to be dominant.
In that case the DM phenomenology is determined by three parameters: $m_N$, $M_{Z'}$ and the gauge coupling $g_{B-L}$.
Figure~\ref{fig:gX} shows contours for the values of the coupling $g_{B-L}$ required in order to successfully reproduce the whole observed DM abundance in the plane $[m_N,\, M_{Z'}]$ (blue thick lines).
The red dotted line corresponds to the resonant annihilation case ($M_{Z'} = 2\, m_N$), and separates the regimes $M_{Z'} \gg 2\, m_N$, where DM annihilates into SM pairs via the $s$-channel exchange of a $Z'$, from the regime  $M_{Z'} \ll 2\, m_N$, where DM annihilates 
into a couple of $Z'$ dominantly via the $t$-channel mediation of $N_R$.
Notice that in the latter limit, also the first 3 diagrams of the second row are relevant, since we could expect $m_{\tilde\sigma'}$ to be of the same order of $M_Z'$.
Nevertheless, achieving the hierarchy $M_{Z'}\ll m_N$ for $g_{B-L}\gtrsim 1$ requires large couplings $y_\sigma$, leading to the non-perturbative regime shown in gray in Fig.~\ref{fig:gX}, and thus this hierarchy is disfavored.
Again, we have performed a full numerical computation with MicrOMEGAs.

In the limit of small momentum transfer, the $Z'$ exchange leads to a spin-independent scattering with nucleons.
The corresponding cross section per nucleon is given by~\cite{Alves:2015pea}
\begin{equation}
    \sigma \simeq \frac{g_{B-L}^4}{\pi} \frac{m_n^2\, m_N^2}{(m_n + m_N)^2\, M_{Z'}^4}\,,
\end{equation}
however, in our numerical analysis we used MicrOMEGAs to determine the values of the spin-independent scattering cross section per nucleon.
PandaX-4T results provide strong bounds to this scenario with large $g_{B-L}$ couplings, constraining spin-independent  cross section up to DM masses of 10~TeV.
Lighter DM masses ($m_N \gtrsim 3$~TeV) are only allowed near the resonance $m_N \simeq 2\, M_{Z'}$.

On the other hand, there are several bounds on the massive gauge boson of $U(1)_{B-L}$~\cite{Escudero:2018fwn}.
For $M_{Z'}$ of few TeV and the large $g_{B-L}$ couplings needed in Fig.~\ref{fig:gX}, the $Z'$ would be copiously produced at the LHC, decaying to electrons and muons 20-30\% of the times, depending if the $Z'$ to the sterile neutrinos channels are kinematically open or not. 
Consequently, LHC searches~\cite{Aad:2019fac, Sirunyan:2021khd} exclude all the viable parameter space for $Z'$ lighter than 6~TeV.
Furthermore, LEP measurements of $e^+e^-\to\ell^+\ell^-$~\cite{LEP:2004xhf} were sensitive to heavy $Z'$ bosons, constraining the mass to be $M_{Z'} > 6\, g_{B-L}$~TeV~\cite{Carena:2004xs}. 
Finally, we note that the required gauge coupling $g_{B-L}$ should be higher than $\sim 0.7$ in order to successfully reproduce the whole observed DM abundance.
Nevertheless, too large $g_{B-L}$ couplings would violate perturbativity, leading to a well-defined allowed area in the parameter space that will be further probed in  future experiments.

\section{Conclusions}\label{Sec:conclusions}

We have constructed a low-scale seesaw model with a local $U(1)_{B-L}$ symmetry, where some of the sterile neutrinos acquire a small Majorana mass radiatively, triggering then an inverse seesaw mechanism to generate tiny masses for the active neutrinos.

In our proposed model, the SM gauge symmetry is supplemented by the spontaneously broken local $U(1)_{B-L}$ and discrete $\mathbb{Z}_4$ symmetries, an extended scalar sector and additional fermionic singlets. 
The latter allows us to cancel the gauge anomalies of $B-L$ and to realize an ISS mechanism for neutrino mass generation. 
The discrete $\mathbb{Z}_4$, spontaneously broken to a residual $\mathbb{Z}_2$, forbids the sterile neutrinos to acquire tree-level Majorana masses, which is instead generated from a one-loop diagram involving the dark sector particles.  
This loop suppression, together with the smallness of the couplings involved in the loop, provides a natural explanation for the smallness of the ISS $\mu$ term. 

Additionally, our model contains both scalar and fermionic DM candidates, whose stability is ensured from the residual $\mathbb{Z}_2$ symmetry. 
We have shown that this model, in addition to providing an origin of the ISS mechanism,  reproduces successfully  the measured value of the DM abundance and is consistent with the constraints arising from the DM direct detection, invisible Higgs decays and LHC $Z'$ searches.
The consistency with all these constraints restricts the mass of the scalar DM candidate to be either $m_h/2$ or larger than about 2~TeV for values of the Higgs portal coupling of order unity. 
On the other hand, in the scenario of fermionic DM candidate, the aforementioned constraints set the masses of the fermionic DM component and the $Z'$ gauge boson to heavier than few TeV, and the $U(1)_{B-L}$ gauge coupling such that $g_{B-L} \gtrsim 0.7$.
Smaller couplings induce a DM overproduction that overclose the Universe.
Alternatively, much higher couplings generate a DM underabundance, compatible with a multicomponent DM scenario.
Future DM searches covering heavy DM candidates, further input from LHC and new detailed measurements of $e^+e^-\to\ell^+\ell^-$ at future lepton colliders will ultimately add new  constraints on the $Z'$ mass and $g_{B-L}$, as well as on the sterile fermion masses, rendering our scenario testable.

\section*{Acknowledgments}
We would like to thank Miguel Escudero for bringing to our attention the LEP constraints on heavy $Z'$ bosons. NB received funding from the Spanish FEDER/MCIU-AEI under grant FPA2017-84543-P, and the Patrimonio Autónomo - Fondo Nacional de Financiamiento para la Ciencia, la Tecnología y la Innovación Francisco José de Caldas (MinCiencias - Colombia) grant 80740-465-2020.
AECH is supported by ANID-Chile FONDECYT 1210378, ANID PIA/APOYO AFB180002 and Milenio-ANID-ICN2019$\_$044. 
AECH is very grateful to the Laboratoire de Physique Théorique, Université Paris-Sud for hospitality and for partially financing his visit where this work was started.
XM is supported by the Alexander von Humboldt Foundation.
This project has received funding/support from the European Union's Horizon 2020 research and innovation programme under the Marie Skłodowska-Curie grant agreement No 860881-HIDDeN, and from the DFG Collaborative Research Center SFB1258. 

\appendix
\section{Anomaly cancellation}
Through the following conditions, we have checked the gauge and  gravitational anomaly cancellation
\begin{align}
    A_{\left[ SU\left( 3\right) _{C}\right] ^{2}U\left( 1\right) _{B-L}}&=2\sum_{i=1}^{3}\left( B-L\right) _{q_{iL}}-\sum_{i=1}^{3}\left[ \left(B-L\right) _{u_{iR}}+\left( B-L\right) _{d_{iR}}\right], \label{anomaly0}\\
    A_{\left[ SU\left( 2\right) _{L}\right] ^{2}U\left( 1\right) _{X}}&=2\sum_{i=1}^{3}\left( B-L\right) _{\ell_{iL}}+6\sum_{i=1}^{3}\left(
B-L\right) _{q_{iL}},  \\
    A_{\left[ U\left( 1\right) _{Y}\right] ^{2}U\left( 1\right) _{B-L}}&=2\sum_{i=1}^{3}\left[ Y_{\ell_{iL}}^{2}\left( B-L\right)_{\ell_{iL}}+3Y_{q_{iL}}^{2}\left( B-L\right) _{q_{iL}}\right]  \notag \\
    &- \sum_{i=1}^{3}\left[ Y_{\ell_{iR}}^{2}\left( B-L\right)_{\ell_{iR}}+3Y_{u_{iR}}^{2}\left( B-L\right) _{u_{iR}}+3Y_{d_{iR}}^{2}\left(B-L\right) _{d_{iR}}\right],\\
    A_{\left[ U\left( 1\right) _{B-L}\right] ^{2}U\left( 1\right) _{Y}}&=2\sum_{i=1}^{3}\left[ Y_{\ell_{iL}}\left( B-L\right)_{\ell_{iL}}^{2}+3Y_{q_{iL}}\left( B-L\right) _{q_{iL}}^{2}\right]  \notag \\
    & -\sum_{i=1}^{3}\left[ Y_{\ell_{iR}}\left( B-L\right)_{\ell_{iR}}^{2}+3Y_{u_{iR}}\left( B-L\right) _{u_{iR}}^{2}+3Y_{d_{iR}}\left(B-L\right) _{d_{iR}}^{2}\right],\\
    A_{\left[ U\left( 1\right) _{B-L}\right] ^{3}}&=\sum_{i=1}^{3}\left[2\left(B-L\right) _{\ell_{iL}}^{3}+6\left( B-L\right) _{q_{iL}}^{3}-3\left(B-L\right) _{u_{iR}}^{3}-3\left( B-L\right) _{d_{iR}}^{3}\right]  \notag \\
    &-\sum_{i=1}^{3}\left[ \left( B-L\right)_{\ell_{iR}}^{3}+\left( B-L\right)_{N_{R_{i}}}^{3}\right] +\sum_{k=1}^{2}\left[ \left( B-L\right) _{\nu_{R_{k}}}^{3}+\left( B-L\right) _{\nu _{S_{k}}}^{3}\right], \\
    A_{\left[ \text{Gravity}\right] ^{2}U\left( 1\right) _{B-L}}&=\,\sum_{i=1}^{3} \left[ 2\left( B-L\right) _{\ell_{iL}}+6\left( B-L\right) _{q_{iL}}-3\left(B-L\right) _{u_{iR}}-3\left( B-L\right) _{d_{iR}}\right]  \notag \\
    & -\sum_{i=1}^{3}\left[ \left( B-L\right) _{\ell_{iR}}+\left( B-L\right)_{N_{R_{i}}}\right] +\sum_{k=1}^{2}\left[ \left( B-L\right) _{\nu_{R_{k}}}+\left( B-L\right) _{\nu _{S_{k}}}\right],  \label{anomaly}
\end{align}
with $A$ the anomaly coefficients appearing in the triangular amplitude involving gauge bosons (corresponding to the groups indicated in the subindices) in the external lines and SM fermions in the internal lines.
Here $B-L$ and $Y$ refer respectively to the charges under $U(1)_{B-L}$ and $U(1)_Y$, as given in Table~\ref{Themodel}.

\section{Stability and unitarity conditions}\label{sec:stability-unitarity}

In order to determine the stability conditions of the scalar potential, one has to analyze its
quartic terms because they will dominate the behavior of the scalar
potential in the region of very large values of the field components. To this end, we introduce the following hermitian bilinear combination of the scalar fields \begin{equation}
a=\phi ^{\dagger }\phi ,\hspace{1cm}b=\sigma ^{\ast }\sigma ,\hspace{1cm}%
c=\zeta ^{\ast }\zeta ,\hspace{1cm}d=\chi ^{2},\hspace{1cm}e=\zeta^{2},
\end{equation}
and rewrite the quartic terms of the scalar potential as follows:
\begin{eqnarray}
V_{4} &=&\left( \sqrt{\lambda }a-\sqrt{\lambda _{\sigma }}b\right)
^{2}+\left( \sqrt{\lambda }a-\sqrt{\lambda _{\zeta }}c\right) ^{2}+\left( 
\sqrt{\lambda }a-\frac{1}{2}\sqrt{\lambda _{\chi }}d\right) ^{2}+\left( 
\sqrt{\lambda _{\sigma }}b-\sqrt{\lambda _{\zeta }}c\right) ^{2}  \notag \\
&&+\left( \sqrt{\lambda _{\sigma }}b-\frac{1}{2}\sqrt{\lambda _{\chi }}%
d\right) ^{2}+\left( \sqrt{\lambda _{\zeta }}c-\frac{1}{2}\sqrt{\lambda
_{\chi }}d\right) ^{2}-2\left( \lambda a^{2}+\lambda _{\sigma }b^{2}+\lambda
_{\zeta }c^{2}\right)   \notag \\
&&+\left( \lambda _{\phi \sigma }+2\sqrt{\lambda \lambda _{\sigma }}\right)
ab+\left( \lambda _{\phi \zeta }+2\sqrt{\lambda \lambda _{\zeta }}\right)
ac+\left( \lambda _{\phi \chi }+\sqrt{\lambda \lambda _{\chi }}\right) ad \\
&&+\left( \lambda _{\zeta \sigma }+2\sqrt{\lambda _{\sigma }\lambda _{\zeta }%
}\right) bc+\left( \lambda _{\zeta \chi }+\sqrt{\lambda _{\zeta }\lambda
_{\chi }}\right) cd+\left( \lambda _{\sigma \chi }+\sqrt{\lambda _{\sigma
}\lambda _{\chi }}\right) bd + \lambda'_{\zeta} \left(e^{2}+h.c\right)\,\notag.
\end{eqnarray}
Following the procedure used for analyzing the stability described in Refs.~\cite{Maniatis:2006fs, Bhattacharyya:2015nca}, we find that our scalar potential will be stable when the following conditions are fulfilled:
\begin{eqnarray}
\lambda \geq 0,\hspace{1cm}&\lambda _{\sigma }&\geq 0,\hspace{1cm}\lambda
_{\zeta }\geq 0,\hspace{1cm}\lambda _{\chi }\geq 0,\hspace{1cm} \lambda'_{\zeta}\geq 0 ,  \notag \\
\lambda _{\phi \sigma }+2\sqrt{\lambda \lambda _{\sigma }} &\geq &0,\hspace{%
1cm}\lambda _{\phi \zeta }+2\sqrt{\lambda \lambda _{\zeta }}\geq 0,\hspace{%
1cm}\lambda _{\phi \chi }+\sqrt{\lambda \lambda _{\chi }}\geq 0,  \notag \\
\lambda _{\zeta \sigma }+2\sqrt{\lambda _{\sigma }\lambda _{\zeta }} &\geq
&0,\hspace{1cm}\lambda _{\zeta \chi }+\sqrt{\lambda _{\zeta }\lambda _{\chi }%
}\geq 0,\hspace{1cm}\lambda _{\sigma \chi }+\sqrt{\lambda _{\sigma }\lambda
_{\chi }}\geq 0.
\end{eqnarray}

Regarding unitarity conditions, they are imposed to avoid that the scattering amplitudes involving physical scalars and longitudinal gauge bosons grow too much with energy. 
In practice, we consider the partial wave expansion, given for an amplitude $\mathcal M$ by
\be
\mathcal M (\theta) = 16\pi\,\sum_{\ell=0}^\infty (2\ell+1)\, a_\ell\, P_\ell(\cos\theta)\,,
\ee
with $P_\ell$ the Legendre polynomials and $a_\ell$ the partial wave of order $\ell$, and  require that the partial waves are bounded from above for any of the possible $2\to2$ scatterings.

This process can be simplified thanks to the equivalence theorem~\cite{Cornwall:1974km, Vayonakis:1976vz, Lee:1977eg, Gounaris:1986cr}, as explained for instance, in Ref.~\cite{Bhattacharyya:2015nca}.
The main idea is that the high-energy behavior of these scatterings can be described using the unphysical scalar particles.
Moreover, these scatterings will be dominated by the dimensionless quartic couplings in Eq.~\eqref{Eq:potential}, which contribute only to the $\ell=0$ partial wave, bounded to be $|a_0|<1$.
Then, we can easily compute the high-energy scattering amplitudes in the unphysical basis, where all the interactions are just given by the quartic couplings, and then obtain the physical unitarity conditions by requiring that the eigenvalues of this $\mathcal S$-matrix are lower than $16\pi$. 

In our scalar sector, we can have two kinds of $2\to2$ scatterings, as the electric charge needs to be conserved. 
From one side, we can have electrically neutral pairs defining the basis 
\be
\left(\phi^-\phi^+,\, \frac{hh}{\sqrt2},\,  \frac{\phi_Z\phi_Z}{\sqrt2},\,  \frac{\tilde\sigma\tilde\sigma}{\sqrt2},\, \frac{\sigma_{Z'}\sigma_{Z'}}{\sqrt2},\, \frac{\tilde\chi\tilde\chi}{\sqrt2},\, \frac{\zeta_R\zeta_R}{\sqrt2},\, \frac{\zeta_I\zeta_I}{\sqrt2},\, \zeta_R\zeta_I\right)\,,
\ee
with the factor $1/\sqrt2$ for identical particles, which leads to the following scattering matrix,
\begin{equation}
\mathcal S_N = 
{\footnotesize
\left(\begin{array}{ccccccccc}
4\lambda & \sqrt2 \lambda & \sqrt2\lambda & \lambda_{\phi\sigma}\sfrac{}{\sqrt2} & \lambda_{\phi\sigma}\sfrac{}{\sqrt2} & \sqrt2\lambda_{\phi\chi}  & \lambda_{\phi\zeta}\sfrac{}{\sqrt2} & \lambda_{\phi\zeta}\sfrac{}{\sqrt2} & 0\\
\sqrt2\lambda & 3 \lambda & \lambda & \lambda_{\phi\sigma}\sfrac{}{2} & \lambda_{\phi\sigma}\sfrac{}{2} &\lambda_{\phi\chi}  & \lambda_{\phi\zeta}\sfrac{}{2} & \lambda_{\phi\zeta}\sfrac{}{2} & 0\\
\sqrt2\lambda &  \lambda & 3\lambda & \lambda_{\phi\sigma}\sfrac{}{2} & \lambda_{\phi\sigma}\sfrac{}{2} &\lambda_{\phi\chi}  & \lambda_{\phi\zeta}\sfrac{}{2} & \lambda_{\phi\zeta}\sfrac{}{2} & 0\\
\lambda_{\phi\sigma}\sfrac{}{\sqrt2} & \lambda_{\phi\sigma}\sfrac{}{2} & \lambda_{\phi\sigma}\sfrac{}{2} & 3\lambda_\sigma &  \lambda_\sigma &  \lambda_{\sigma\chi} & \lambda_{\zeta\sigma}\sfrac{}{2} & \lambda_{\zeta\sigma}\sfrac{}{2} & 0\\
\lambda_{\phi\sigma}\sfrac{}{\sqrt2} & \lambda_{\phi\sigma}\sfrac{}{2} & \lambda_{\phi\sigma}\sfrac{}{2} & \lambda_\sigma &  3\lambda_\sigma &  \lambda_{\sigma\chi} & \lambda_{\zeta\sigma}\sfrac{}{2} & \lambda_{\zeta\sigma}\sfrac{}{2} & 0\\
\sqrt2 \lambda_{\phi\chi} &  \lambda_{\phi\chi} &  \lambda_{\phi\chi} &  \lambda_{\sigma\chi} &  \lambda_{\sigma\chi} & 3 \lambda_\chi &  \lambda_{\zeta\chi} &  \lambda_{\zeta\chi}  & 0\\
\lambda_{\phi\zeta}\sfrac{}{\sqrt2} & \lambda_{\phi\zeta}\sfrac{}{2} & \lambda_{\phi\zeta}\sfrac{}{2} & \lambda_{\zeta\sigma}\sfrac{}{2} & \lambda_{\zeta\sigma}\sfrac{}{2} &  \lambda_{\zeta\chi} & 3\lambda_\zeta +6\lambda'_\zeta &  \lambda_\zeta -6\lambda'_\zeta  & 0 \\
\lambda_{\phi\zeta}\sfrac{}{\sqrt2} & \lambda_{\phi\zeta}\sfrac{}{2} & \lambda_{\phi\zeta}\sfrac{}{2} & \lambda_{\zeta\sigma}\sfrac{}{2} & \lambda_{\zeta\sigma}\sfrac{}{2} &  \lambda_{\zeta\chi} &  \lambda_\zeta -6\lambda'_\zeta &  3\lambda_\zeta +6\lambda'_\zeta  & 0 \\
0 & 0 & 0 & 0 & 0 & 0 & 0 & 0 & 2\lambda_\zeta -12\lambda'_\zeta
\end{array}\right)
}.
\end{equation}
From the other side, considering electrically charged pairs and taking the basis $(h\phi^+,\phi_Z\phi^+, \tilde\sigma\phi^+,\\ \sigma_{Z'}\phi^+, \tilde\chi\phi^+, \zeta_R\phi^+, \zeta_I\phi^+)$, we have
\begin{equation}
    \mathcal S_C = {\rm diag}\left( 2 \lambda,\, 2 \lambda,\, \lambda_{\phi\sigma},\, \lambda_{\phi\sigma},\, 2 \lambda_{\phi\chi},\, \lambda_{\phi\zeta},\, \lambda_{\phi\zeta}\right).
\end{equation}
The unitarity conditions imply that the modulus of the eigenvalues of $\mathcal S_N$ and $\mathcal S_C$ lie below $16\pi$.

\bibliographystyle{utphys}
\bibliography{Refs}
\end{document}